\documentclass{JHEP3}
\usepackage{amsmath,amssymb,amsthm,dcpic,pictexwd,graphicx}
\usepackage[mathscr]{eucal}

\numberwithin{equation}{section}

\def\defeq{\buildrel\rm def\over=}

\def\HH{{H\!H}}

\def\BC{\mathbb{C}}
\def\BR{\mathbb{R}}

\newcommand{\MR}[1]{{\mathbb{R}^{#1}}}            % Real numbers
\newcommand{\MC}[1]{{\mathbb{C}^{#1}}}            % Complex numbers
\newtheorem{defn}{Definition}
\newtheorem{thm}{Theorem}
\newtheorem*{ithm}{Theorem}
\newtheorem{lma}{Lemma}

\def\Hom{\mathrm{Hom}}
\def\End{\mathrm{End}}

\def\Obj{\mathrm{Obj}}
\def\Arr{\mathrm{Arr}}

\def\Ext{\mathrm{Ext}}

\def\dim{\mathrm{dim}}
\def\mcO{\mathcal{O}}
\def\mcA{\mathcal{A}}

\def\mcC{\mathcal{C}}
\def\mcD{\mathcal{D}}
\def\mcE{\mathcal{E}}
\def\mcF{\mathcal{F}}

\def\mcP{\mathcal{P}}
\def\id{\mathrm{id}}

\def\ie{\textit{i.e.,\ }}

\title{Stability Conditions and Branes at Singularities}

\author{Aaron Bergman\\ George P. \& Cynthia W. Mitchell Institute for Fundamental Physics\\
             Texas A\&M University\\
             College Station, TX 77843-4242\\ {~}\\
             E-mail: \email{abergman@physics.tamu.edu}}

\preprint{MIFP--07--05\\\hepth{0702092}}

\abstract{I use Bridgeland's definition of a stability condition on a triangulated category to investigate the stability of D-branes on Calabi-Yau cones given by the canonical line bundle over a del Pezzo surface. In this context, I prove the existence of the decay of a D3-brane into a set of fractional branes. This is an important aspect of the derivation of quiver gauge theories from branes at singularities via the technique of equivalences of categories. Some important technical aspects of this equivalence are discussed. I also prove that the representations corresponding to skyscraper sheaves supported off the zero section are simple.}

\begin{document}

\section{Introduction}\label{sec:intro}

The use of the topological B-model to explore the gauge theories living on D-branes at singularities has been extremely fruitful in recent years \cite{Wijnholt:2002qz,Herzog:2003dj,Herzog:2004qw,Aspinwall:2004vm,BridgeTStruct,Bergman:2006gv,Bergman:2005kv,Bergman:2005ba}. The fundamental feature of this technique is an equivalence of categories between the derived category of coherent sheaves on the resolved singularity and the derived category of representations of the algebra described by the quiver in the dual gauge theory. However, because of the reliance on the topological B-model, this technique can only describe features related to the complex geometry, \ie the F-terms in the gauge theory. To go further, we must bring in the K\"ahler geometry of the resolved singularity.

One important consequence of the K\"ahler geometry is towards the stability of D-branes. In particular, the geometry determines the central charge, the phase of which is related to fractional part of the grading of the brane. Through $\pi$-stability as formulated by Douglas \cite{Douglas:2000gi} and its generalization due to Bridgeland \cite{Bridgeland:2002sc}, this grading determines the topological D-branes that remain stable when passing to the physical string. The difference in gradings between two branes also determines the masses of the string modes stretching between them.

Understanding stability is important because we need to know that a D3-brane located at the tip of the cone is marginally unstable to decay to a particular set of fractional branes. The strings between these fractional branes then determine the quiver gauge theory. In particular, we need to show both that this decay exists and that there exists a point in the K\"ahler moduli space where all the gradings align. Since the computation of gradings in terms of geometry in type IIB string theory involves unknown corrections, this is usually accomplished by passing to the mirror where the grading is related to the phase of the period of the holomorphic three-form over the relevant three-cycle. By solving the Picard-Fuchs equations, one can then determine how the gradings behave as we move around in moduli space. This has been accomplished in many examples \cite{Aspinwall:2004vm,Aspinwall:2004jr} but can be difficult in general.

In this paper, I will avoid this difficulty by simply postulating the existence of the needed set of gradings. In particular, Bridgeland's notion of a stability condition on a triangulated category reproduces the features of $\pi$-stability and does not refer directly to the metric on our variety. One can then consider the space of all such stability conditions. This is \textit{not} the K\"ahler moduli space of the relevant variety; it is much too big. It may be a generalization of the K\"ahler moduli space of the Calabi-Yau (beyond the usual generalization defined to be the complex structure moduli space of the mirror.) On the other hand, it may be that Bridgeland's definition is too general, and we need further conditions to reproduce the physical moduli space. I will not address this issue here except to note (following Bridgeland) that there ought to exist a generalization of the complex structure moduli space whose tangent space is given by the full Hochschild cohomology of the variety. The possibility of such an extended moduli space is also mentioned by Witten \cite{Witten:1991zz}.

Leaving these issues behind, Bridgeland's stability conditions are defined as a structure on a trangulated category without any reference to the Abelian category from whence it came. This makes it ideal to apply to our equivalence of categories. Unfortunately, due to the noncompactness of our geometry, there are a number of technical issues relating the existence of infinite dimensional cohomology groups. Bridgeland avoids this issue by restricting his attention to sheaves that are supported on the `base' of the resolved geometry. This is not sufficient for our application as we wish to reproduce the entire cone as a moduli space of stable representations. In this paper, I will show how to define a stability condition that reproduces a formulation of GIT-stability due to King \cite{King:1994mr} (often called $\theta$-stability in the physics literature) and which reproduces the moment map used in the symplectic reduction construction of SUSY gauge theory moduli spaces.

With this in hand, we can make rigorous a number of results in the physics literature including a demonstration that the needed D3-brane decay exists (we have, of course, postulated the needed alignment of the gradings rendering that no longer an issue). In addition, a result of Bridgleland \cite{Bridgeland:2006ss} tells us that moving around in the space of stability conditions is often related to a procedure called tilting. This tilting is related to Seiberg duality as was first demonstrated by Berenstein and Douglas \cite{Berenstein:2002fi} and further explored by Herzog \cite{Herzog:2004qw} and Aspinwall \cite{Aspinwall:2004vm}. As an example of this, the study of stability conditions on ALE singularities \cite{Bridgeland:2005sk} reproduces the beautiful picture of Cachazo \textit{et al} \cite{Cachazo:2001sg} where Seiberg dualities arise as one approaches the walls of the Weyl chamber that is the K\"ahler moduli space. I have little to add to this story, so I will not discuss it further. Stability of D-branes on cones is also discussed from a more physical perspective in \cite{Aspinwall:2004mb}.

The main mathematical result in this paper is the following theorem:

\begin{ithm} Let $Y$ be a smooth variety with a locally free tilting sheaf $T$,  $\pi$ be the projection from $Y$ to its affinization, and $p$ be a point in $Y$ such that $\pi^{-1}\pi(p) = p$. Then the $\End_Y(T)$ module corresponding to $\mcO_p$ is  a simple module.\end{ithm}

\noindent Recall that the affinization of a variety is the spectrum of its ring of global functions. The physical content of this result is as follows. The singularities we will deal with are collapsed del Pezzo surfaces in a Calabi-Yau. Thus, a local model for the singularity will be $K_X$, the total space of the canonical line bundle over a del Pezzo surface $X$. In string theory compactified on $\MR{3,1} \times K_X$, the D3-brane fills the $\MR{3,1}$ and is thus represented by a skyscraper sheaf on $K_X$. This theorem tells us that when the D3-brane is off the zero section of $K_X$, \ie away from the `singularity', it is stable for any stability condition associated to a quiver representation. This includes all the stability conditions defined in this paper.

This paper is organized as follows. In section two, I will give a lightning review of the equivalences of categories that are at the heart of this construction paying attention to some technical points that often are elided. In section three, I will discuss the stability of topological D-branes and Bridgeland's formalization thereof. In section four, I will discuss Bridgeland's results on the construction of stability conditions from t-structures. In section five, these techniques will be applied to our situation. I conclude in section six with a proof of the above theorem.

\section{Equivalences of categories}\label{sec:cat}

This section will be a brief overview of the geometry and the equivalence of categories that we will utilize. For a more complete presentation of my view on the subject, please see \cite{Bergman:2006gv}. The original references are given in the introduction.

Let $X$ be a del Pezzo surface and let $K_X$ be the total space of the canonical line bundle over $X$. We will denote the projection $\pi : K_X \to X$ and the zero section $s : X \to K_X$. It is straightforward to see that the canonical bundle over $K_X$ is trivial. We can place a Calabi-Yau metric on $K_X$. By collapsing the zero section of $K_X$, we give rise to a singular geometry that is a local model for a collapsing del Pezzo surface inside a Calabi-Yau. We will work with the resolved geometry, \ie the line bundle $K_X$. It follows from a theorem of Bridgeland \cite{Bridgeland:2002fd} that all crepant resolutions of a singularity in a CY 3-fold have equivalent derived categories, so by working solely in that context we do not lose any generality by our choice of resolution.

A D3-brane filling the transverse $\MR{3,1}$ is located at a point on the CY cone, $K_X$, and is represented by a skyscraper sheaf. We will argue later that when the D3-brane is located along the zero section of $K_X$, it will be unstable to decay to a set of ``fractional branes". The string states between these fractional branes will give rise to the quiver gauge theory. The starting point for this paper will be that there is an equivalence between the derived category of representations of this quiver and the derived category of coherent sheaves on $K_X$. To construct such an equivalence, we begin with an exceptional collection on $X$. This is a collection of coherent sheaves $E_i$ on $X$ which generate the derived category and satisfy
\begin{eqnarray}
\label{exccond1}
\Ext_X^i(E_a,E_b) &=& 0 \text{\quad for } i \neq 0\ , \\
\label{excond2}
\Hom_X(E_a,E_b) &= &0 \text{\quad for } a > b\ , \\
\label{excond3}
\Hom_X(E_a,E_a) &=& \BC \ .
\end{eqnarray}
The direct sum $T = \bigoplus E_i$ is a tilting object on X, and it follows from a theorem of Rickard \cite{Rickard:1988mt} that there is an equivalence of categories between the bounded derived categories of coherent sheaves on $X$ and finite dimensional modules over the algebra $\End_X(T)$. The construction of a quiver algebra from this data was first done in \cite{BondalQuiv}.

We want an equivalence of categories for $K_X$, however, and not just $X$. We can accomplish this (following Bridgeland \cite{BridgeTStruct}) by imposing the further condition that
\begin{equation}
\Ext^i_X(E_a,E_b\otimes \omega_X^p) = 0 \quad \mbox{for } i \neq 0, p \le 0\ .
\end{equation}
Then  $\pi^*T = \bigoplus \pi^*E_i$ generates the derived category of coherent sheaves on $K_X$ and is a tilting object. Because $K_X$ is not projective, we must be careful in the precise statement of the results of Rickard's theorem. We have an equivalence of categories between the \textit{unbounded} derived category of quasicoherent sheaves on $K_X$ and the unbounded derived category of modules of $A \defeq \End_{K_X}(\pi^*T)^\mathrm{op}$. (By taking the opposite algebra, we interchange left- and right-modules.) This restricts to an equivalence of full categories between the subcategories of objects isomorphic to \textit{perfect} objects. For the case of modules over $A$, these are bounded complexes of finitely-generated projective $A$-modules. As $A$ is noetherian and of finite global dimension, $\mbox{per }A \cong \mcD^b(A\mathrm{- fgmod})$, the bounded derived category of finitely generated $A$-modules. On $K_X$, the perfect complexes are those locally isomorphic to bounded complexes of vector bundles. As $K_X$ is smooth, any coherent sheaf can be resolved\footnote{There is a subtlety here in that this is true algebraically, but not necessarily analytically. However, because this is the total space of a negative line bundle over a projective variety, the needed resolutions exist (assertion 3, p. 701 \cite{GriHar}).} in terms of vector bundles, so we have $\mbox{per }K_X \cong \mcD^b(\mathrm{Coh}(K_X))$. Thus, Rickard's theorem gives us the equivalence of categories $ \mcD^b(\mathrm{Coh}(K_X)) \cong \mcD^b(A\mathrm{- fgmod})$.

The difficulty with this nonprojective case is that the algebra, $A$, is infinite dimensional. Thus, \textit{finitely generated} modules are not the same as \textit{finite dimensional} modules. We will denote the latter category $A \mathrm{- fdmod}$. The category $\mcD^b(A \mathrm{- fdmod})$ is a full subcategory of $\mcD^b(A \mathrm{- fgmod})$. It would be very interesting to understand what the corresponding full subcategory of $\mcD^b(\mbox{Coh}(K_X))$ is. It certainly contains, for example, the category $\mcD^b(\mbox{Coh}_c(K_X))$ consisting of complexes whose constituent sheaves have compact support. It also is worth considering $\mcD^b_\mathrm{fd}(A \mathrm{- fgmod})$, the full subcategory of $\mcD^b(A \mathrm{- fgmod})$ consisting of objects whose cohomology modules are finite-dimensional. Similarly, one can consider $\mcD^b_c(\mathrm{Coh(K_X)})$, the full subcategory of $\mcD^b(\mathrm{Coh(K_X)})$ consisting of objects whose cohomology sheaves have compact support. This latter category is, in fact, a Calabi-Yau category of dimension three. It is proven in \cite{Bergman:2008yi} that
\begin{equation}
\mcD^b_c(\mathrm{Coh(K_X)}) \cong \mcD^b_\mathrm{fd}(A \mathrm{- fgmod})\ .
\end{equation}

As was the case with $\End_X(T)$, the algebra $A$ can be interpreted as the path algebra of a quiver with relations. In this case, the quiver has loops, reflecting the infinite dimensionality of $A$. For a discussion of the construction of this quiver, please see \cite{Herzog:2004qw,Aspinwall:2004vm,Bergman:2006gv,Bergman:2005ba}. As discussed therein, we have the correspondence $\pi^*E_i \leftrightarrow P_i$ where $P_i$ is the projective representation associated to the node $i$ of the quiver. Note that the $P_i$ are infinite dimensional but finitely generated reflecting the noncompactness of the support of the $\pi^*E_i$. We also have the representations $S_i$ defined by $S_i(j) = \MC{\delta_{ij}}$ with all arrows given by the zero map. These are simple representations and correspond to complexes of coherent sheaves whose cohomology objects are supported on the zero section of $K_X$. Note that these are \textit{not} the only simple objects in the abelian category of modules even if we impose the finite generation or finite dimensionality condition.

This motivated Bridgeland to introduce yet another category \cite{Bridgeland:2006sc}. Let $\mcD_0^b(A\mathrm{- fgmod})$ be the smallest full subcategory of $\mcD^b(A\mathrm{- fgmod})$ containing the objects $S_i$. There is a corresponding category $\mcD_0^b(\mathrm{Coh}(K_X))$ which is a full subcategory of $\mcD^b(\mathrm{Coh}(K_X))$. We will call an $A$-module \textit{tiny} if it is an object in the smallest extension-closed subcategory of $A \mathrm{- mod}$ containing the modules $S_i$. Then $\mcD_0^b(A\mathrm{- fgmod})$ can also be characterized as the full subcategory of $\mcD^b(A\mathrm{- fgmod})$ consisting of objects whose cohomology modules are tiny. Since the simple modules are supported on the zero section, it turns out that $\mcD_0^b(\mathrm{Coh}(K_X))$ can be characterized as the full subcategory of  $\mcD^b(\mathrm{Coh}(K_X))$ consisting of objects whose cohomology sheaves are supported on the zero section of $K_X$. By construction, we have the equivalence of categories $\mcD_0^b(\mathrm{Coh}(K_X)) \cong \mcD_0^b(A\mathrm{- fgmod})$.

An important property of the abelian category of tiny modules is that it is of \textit{finite length} with simple objects $S_i$. This means that for any module $M$, there is a finite sequence of submodules
\begin{equation}
0 = F_0 \subset F_1 \subset F_2 \subset \dots \subset F_{n-1} \subset F_n = M\
\end{equation}
such that each quotient $F_i/F_{i-1}$ is one of the $S_i$. We will later use the existence of such a sequence to show that, when the gradings align, the D3-brane becomes marginally stable against decay to fractional branes (which we recall are represented by the $S_i$) precisely when it is located on the zero section of $K_X$.

\section{Stability of topological D-branes}\label{sec:stab}

BPS D-branes have associated to them a central charge. For special lagrangian A-branes this is given by the integral of the holomorphic three-form pulled back to the worldvolume of the branes. In the B-model, no exact formula is known, but the leading term is
\begin{equation}
\label{cencha}
Z(E) = \int_M e^{B + iJ} ch(E)\sqrt{td(M)} + \dots
\end{equation}
where $M$ is the Calabi-Yau, $B$ is the usual B-field and $J$ is the K\"ahler form. For a CY 3-fold, there are no perturbative corrections and the nonperturbative corrections are given by a power series in $q_i = \exp{2\pi i \left(\int_{C_i} B+iJ\right)}$ with no constant term \cite{Aspinwall:2004jr}.

The phase of this central charge is proportional to the grading of the D-brane. More precisely, the grading of a brane $E$ is given by
\begin{equation}
\label{cenpha}
\xi(E) = \frac{1}{\pi}\arg Z(E)\ .
\end{equation}
This is incomplete, however. Douglas has argued \cite{Douglas:2000gi} that we should define a real valued grading such that its reduction modulo 2 is given by \eqref{cenpha}. Furthermore, we require that
\begin{equation}
\label{censhift}
\xi(E[n]) = \xi(E) + n\ .
\end{equation}
In fact, it will best to only call this a `grading' when the object $E$ is stable and to not assign a grading to unstable branes.

Recall that the spectrum of string states between two topological D-branes, $E$ and $F$, are given by the groups $\Hom(E,F[i])$ where $i$ is the level of the state. Douglas showed \cite{Douglas:2000gi} that these correspond to physical string states with mass
\begin{equation}
\label{strmass}
m^2 = \frac{1}{\alpha'}\left(i-1 + \xi(E) - \xi(F)\right)\ .
\end{equation}
Notice the compatibility with the relation \eqref{censhift}.

The starting point for Douglas's notion of stability is that a pair of D-branes will bind if there is a tachyonic string between them. Thus, let $E$ and $F$ be objects in the derived category and $f \in \Hom(E,F)$ a string between them. The mass of this string is given by \eqref{strmass} with $i=0$. Thus, if $\xi(E) - \xi(F) < 1$, the D-branes will form a bound state. This bound state will be isomorphic to Cone$(E\to F)$. A D-brane will be considered stable if it is not isomorphic to the cone of a non-tachyonic map between any two \textit{stable} branes. In other words, an object $E$ is stable if there are no distinguished triangles
\begin{equation}
\label{decaytri}
\begin{split}
\begindc{\commdiag}[3]
\obj(20,25)[e]{$E$}
\obj(10,10)[a1]{$A_1$}
\obj(30,10)[a2]{$A_2$}
\mor{a1}{e}{}
\mor{e}{a2}{}
\mor{a2}{a1}{}[\atright,\dasharrow]
\enddc
\end{split}
\end{equation}
with $A_1$ and $A_2$ stable with gradings that obey $\xi(A_1) > \xi(A_2)$. Here (and henceforth), the dashed line represents a map from $A_2$ to $A_1[1]$. Thus, it represents a state with mass given by \eqref{strmass} as $m^2 = \frac{1}{\alpha'}(\xi(A_1) - \xi(A_2)) > 0$. The existence of such a triangle means that the objects $A_1$ and $A_2$ do not bind to form $E$ and instead destabilize it.

This condition is very difficult to deal with in practice. It is also deficient in that it only considers two-body decays. Bridgeland invented a formalization and generalization of this notion which I will now describe\footnote{The remainder of this section closely follows \cite{Bridgeland:2002sc}.}. The first ingredient we will need is the notion of the central charge. Since we are working with an arbitrary triangulated category, we should not refer to things like integrals as in \eqref{cencha}. We recognize the combination of the Chern and Todd classes as giving a map from the Grothendieck K-theory of the derived category of coherent sheaves to ordinary cohomology. The equation \eqref{cencha} then gives a map to $\BC$. Thus, we define the central charge for an arbitrary triangulated category $\mcD$ to be a map $Z : K(\mcD) \to \BC$. The group $K(\mcD)$ can be infinite dimensional, however (as in the case of a torus), so it may be worthwhile to replicate the feature of \eqref{cencha} where we first apply a Chern character. One possibility is that we define the central charge to be a map from $\HH_0(\mcD)$ to $\BC$. (Bridgeland suggests using periodic cyclic cohomology.)

The next ingredient we need is a notion of the semistable objects of a given grade. This gives rise to the notion of a slicing defined as follows.
\begin{defn} A \textit{slicing} $\mcP$ of a triangulated category $\mcD$ consists of full additive subcategories $\mcP(\xi)$ of $\mcD$ for each $\xi \in \BR$ satisfying the following axioms:
\begin{enumerate}
\item for all $\xi \in \BR$, $\mcP(\xi+1) = \mcP(\xi)[1]$.
\item if $\xi_1 > \xi_2$ and $A_j \in \mcP(\xi_j)$, then $\Hom(A_1,A_2) = 0$.
\item for each nonzero object $E \in \mcD$, there exists a finite sequence of real numbers $\xi_1 > \xi_2 > \dots > \xi_n$ and a collection of triangles
\begin{equation}
\begin{split}
\begindc{\commdiag}[3]
\obj(3,25)[0]{$0=$}
\obj(10,25)[e0]{$E_0$}
\obj(32,25)[e1]{$E_1$}
\obj(54,25)[e2]{$E_2$}
\obj(76,25)[ed]{$\dots$}
\obj(98,25)[en1]{$E_{n-1}$}
\obj(120,25)[en]{$E_n$}
\obj(127,25)[e]{$=E$}
\obj(21,10)[a1]{$A_1$}
\obj(43,10)[a2]{$A_2$}
\obj(109,10)[an]{$A_n$}
\mor{a1}{e0}{}[\atright,\dasharrow]
\mor{e0}{e1}{}
\mor{e1}{a1}{}
\mor{e1}{e2}{}
\mor{e2}{a2}{}
\mor{a2}{e1}{}[\atright,\dasharrow]
\mor{e2}{ed}{}
\mor{ed}{en1}{}
\mor{en1}{en}{}
\mor{en}{an}{}
\mor{an}{en1}{}[\atright,\dasharrow]
\enddc
\end{split}
\end{equation}
with $A_j \in \mcP(\phi_j)$ for all $j$.
\end{enumerate}
\end{defn}

The physical import of these axioms is as follows. Each subcategory is the category of semistable objects of a fixed grading. That the subcategories are additive encodes the idea that the direct sum of two objects with the same grading has the same grading (and is marginally stable, so we can assign a grading). The first axiom is a restatement of equation \eqref{censhift}. The second axiom is needed to ensure that the relevant objects are in fact semistable. Finally, the third axiom encodes the decay of any object $E$ into a finite set of semistable objects $A_i$. The two body decays considered above constitutes the case $n=2$, and the decay chain simplifies to the triangle \eqref{decaytri}.

Finally, we need a compatibility of the slicing and the central charge.
\begin{defn}
A stability condition on a triangulated category $\mcD$ is given by a slicing of the category and a central charge $Z : K(\mcD) \to \BC$ such that for $E \in \mcP(\xi)$
\begin{equation}
Z(E) = m(E)e^{i\pi \xi} \quad \mbox{with } m(E) \in \BR^{>0}\ .
\end{equation}
\end{defn}

\noindent It turns out that the categories $\mcP(\xi)$ are, in fact, abelian. As mentioned above, the semistable branes are the objects in $\mcP(\xi)$ for some $\xi$, and now we can add that the stable objects are precisely the simple semistable objects considered in their respective abelian subcategory.

Bridgeland has shown that, after adding a technical condition, one can form nice moduli spaces of stability conditions. In particular, the local deformations of a stability condition are precisely the deformations of the central charge. In other words, the space of infinitesimal deformations is the dual vector space to $K(\mcD)$. However, this can be infinite dimensional and is certainly not the same as $H^{1,1}(M)$. Even if we restrict to $\HH_0(\mcD)$ as above, the HKR theorem tells us that $\HH_0(\mcD(\mathrm{Coh}(X))) = \bigoplus H^i(X,\Omega^iX)$. Nonetheless, we will proceed assuming that the stability conditions we define are, in fact, physical.

\section{Constructing stability conditions}\label{sec:con}

The goal of the next two sections is to construct stability conditions related to the space $K_X$ discussed in section \ref{sec:cat}. We have a plethora of triangulated categories to choose from. Before discussing the  issues with each choice, let us see how one can construct a stability condition. The first tool we will need is that of a t-structure. This is a tool for finding an abelian category inside of a triangulated category. To motivate the definition, consider the case of the derived category of an abelian category. We can see that the objects of the original category are exactly the length one complexes located at position zero. To make this more formal, we use the fact that we can associate cohomology objects to any object in a derived category. Define $\mcD^{\ge n}$ be the full subcategory of $\mcD$ of objects, $K$, such that $H^i(K) = 0$ for $i<n$. Note that $\mcD^{\ge n} = \mcD^{\ge 0}[-n]$. One can similarly define $\mcD^{\le n}$.  Then $\mcD^{\ge 0} \cap \mcD^{\le 0}$ is exactly the abelian category we began with. This can be formalized as follows \cite{Bridgeland:2002sc}.
\begin{defn}
A t-structure on a triangulated category $\mcD$ is a pair of strictly full subcategories $\mcD^{\le 0}$ and $\mcD^{\ge 0}$ such that
\begin{enumerate}
\item $\mcD^{\le 0} \subset \mcD^{\le 1}$ and $\mcD^{\ge 1} \subset \mcD^{\ge 0}$
\item $\Hom(X,Y) = 0$ for $X \in \Obj(\mcD^{\le 0})$ and $Y \in \Obj(\mcD^{\ge 1})$
\item For any $X\in\Obj(\mcD)$, there exists a distinguished triangle $A \to X \to B \to A[1]$ with $A \in \Obj(\mcD^{\le 0})$ and $B \in \Obj(\mcD^{\ge 1})$
\end{enumerate}
where $\mcD^{\ge n} = \mcD^{\ge 0}[-n]$ and $\mcD^{\le n} = \mcD{\le 0}[-n]$.
\end{defn}

$\mcD^{\ge 0} \cap \mcD^{\le 0}$ is called the \textit{heart} (or core) of the t-structure, and it is a nontrivial theorem that it is an abelian category. It is not necessarily true, however, that the derived category of the core of a t-structure is the original triangulated category, although this is obviously true for the case where the t-structure reflects that our triangulated category is the derived category of an abelian category. Regardless, the t-structure allows us to define cohomology functors valued in the heart. A t-structure is called bounded if $\cap_{i=0}^\infty \mcD^{\ge n} = \cap_{i=0}^\infty \mcD^{\le n} = 0$ and there are only a finite number of nonzero cohomology objects for any object in $\mcD$. 

Given a slicing as defined above and an interval $I \subset \BR$, we can define $\mcP(I)$ to be the extension closed subcategory generated by the $\mcP(\xi)$ for all $\xi \in I$. Then, it is shown in \cite{Bridgeland:2002sc} that, for any $\xi$, there exists a t-structure with core $\mcP((\xi,\xi+1])$.

The way we will define a stability condition on a triangulated category is to first choose a bounded t-structure and then define a notion of stability on the associated abelian category which we will denote $\mcA$. For that, we need a central charge and an associated decomposition into semistable objects. A central charge is defined similarly to the central charge on the triangulated category: it is a function $Z : K(\mcA) \to \BC$ such that, for all nonzero objects in $E \in \mcA$, $Z(E) = r e^{i\pi\xi}$ with $r>0$ and $\xi \in (0,1]$. Given a central charge, we can define a semistable object:
\begin{defn}
\label{stabdef}
An object $E \in \mcA$ is said to be \textbf{semistable} with respect $Z : K(\mcA) \to \BC$ if every nontrivial subobject $F \subset E$ satisfies $\xi(F) \le \xi(E)$.
\end{defn}

Next, we introduce the decomposition into semistable objects:

\begin{defn}
A central charge (or stability function in Bridgeland's terminology) has the \textbf{Harder-Narasimhan property} if, for every $E$ a non-zero object of $\mcA$, there exists a sequence of objects $E_i$ such that
\begin{equation}
0= E_0 \subset E_1 \subset \dots \subset E_{n-1} \subset E_n = E
\end{equation}
where the quotients $F_i = E_i / E_{i-1}$ are semistable and
\begin{equation}
\xi(F_1) > \xi(F_2) > \dots > \xi(F_{n-1}) > \xi(F_n)\ .
\end{equation}
\end{defn}

Now, Proposition 5.3 of \cite{Bridgeland:2002sc} states that a stability condition on a triangulated category $\mcD$ is equivalent to giving a bounded t-structure on $\mcD$ and a central charge obeying the Harder-Narasimhan property on its heart. The intuition for this result is as follows. The central charge on the heart of the t-structure allows us to define additive subcategories $\mcP(\xi)$ for $\xi \in (0,1]$. However, by property 1 in the definition of a slicing, this determines the $\mcP(\xi)$ for all $\xi \in \BR$. The decompositions from the Harder-Narasimhan property then fit together to give the decompositions in the slicing. The fact that this theorem goes the other way is also important. Given a stability condition, $\mcP((0,1])$ is the heart of a t-structure, and the central charge restricted to that subcategory has the Harder-Narasimhan property. In other words, we can determine the set of semistable objects solely be examining that abelian category.

Finally, we note that the Harder-Narasimhan property is not a particularly stringent condition on the central charge given the following result (Prop 2.4 of \cite{Bridgeland:2002sc}):

\begin{thm}
\label{hnthm}
Suppose $\mcA$ is an abelian category with a central charge $Z : K(\mcA) \to \BC$ satisfying the chain conditions
\begin{enumerate}
\item there are no infinite sequences of subobjects in $\mcA$
\begin{equation}
\dots \subset E_{j+1} \subset E_j \subset \dots \subset E_2 \subset E_1
\end{equation}
with $\xi(E_{j+1}) > \xi(E_j)$ for all $j$.
\item there are no infinite sequences of quotients in $\mcA$
\begin{equation}
E_1 \to E_2 \to \dots \to E_j \to E_{j+1} \to \dots
\end{equation}
with $\xi(E_j) > \xi(E_{j+1})$ for all $j$.
\end{enumerate}
Then $Z$ has the Harder-Naramsimhan property.
\end{thm}

\section{Stability conditions associated to quivers}\label{sec:quiver}

We now return to the categories introduced in section \ref{sec:cat}. Recall that these were $\mcD^b(A\mathrm{- fgmod})$, the bounded derived category of finitely generated $A$-modules, $\mcD^b(A\mathrm{- fdmod})$, the bounded derived category of finite dimensional $A$-modules, $\mcD^b_\mathrm{fd}(A\mathrm{- fgmod})$, the full subcategory of $\mcD^b(A\mathrm{- fgmod})$ whose cohomology modules are finite dimensional and $\mcD_0^b(A\mathrm{- fgmod})$, the full subcategory of $\mcD^b(A\mathrm{- fgmod})$ whose cohomology modules are tiny and which is equivalent to the full subcategory of $\mcD^b(\mathrm{Coh}(K_X))$ whose cohomology sheaves supported on the zero section of $K_X$. Each has a t-structure associated with being a derived category. Thus, we need a central charge with the Harder-Narasimhan property in order to place a stability condition on them. Each of these categories has advantages and disadvantages towards that end. In particular $\mcD^b(A\mathrm{- fgmod})$ has a particularly simple K-theory: it is a vector space generated by the representatives of the exceptional collection. Unfortunately, because many of the modules are infinite dimensional, it is by no means clear that the conditions of theorem \ref{hnthm} hold.

On the other hand, the theorem is obvious for $A\mathrm{- fdmod}$ on dimensional grounds. In this case, however, the K-theory may be quite complicated. In part because of these dueling difficulties, Bridgeland chose to work with $\mcD_0^b(A\mathrm{- fgmod})$. Its heart is, by construction, a finite length category, and theorem \ref{hnthm} is again obvious. In addition, because the $S_i$ are the only simple objects, the K-theory is a vector space generated by the classes of the $S_i$. Thus, we can give a stability condition on this category by assigning a number in the upper-half plane for each simple object $S_i$. Nonetheless, from the point of view of the physics, this category is unsatisfactory: it only describes branes supported at the zero section of $K_X$. The gauge theory clearly describes more than that, however. For example, it is shown in \cite{Bergman:2005kv,Bergman:2005mz} that the moduli space of vacua of the quiver gauge theory is precisely the singular cone, and that when the FI-terms are turned on, one partially or completely desingularizes the cone. We would like to understand this result in the context of this paper.

While I do not understand the K-theory of the abelian category $A\mathrm{- fdmod}$, there does exist a map $K(A\mathrm{- fdmod}) \to \mathbb{N}^\text{\#nodes}$ given by the dimension vector of the representation. This can be seen by noting that the dimension vectors in a short exact sequence obey precisely the same relation as that which defines the Grothendieck group. We can then define a central charge by $Z(S_i) = r_i e^{i\pi\xi_i}$ for $r_i \ge 0$ and $0 < \xi_i \le 1$. By the theorem, this central charge satisfies the Harder-Narasimhan property and thus defines a stability condition on the category $\mcD^b(A\mathrm{- fdmod})$. In addition, since the heart of the standard t-structure on $\mcD^b(A\mathrm{- fgmod})$ when restricted to $\mcD^b_\mathrm{fd}(A\mathrm{- fgmod})$ is also $A\mathrm{- fdmod}$, this also defines a stability condition on $\mcD^b_\mathrm{fd}(A\mathrm{- fgmod})$.\footnote{Note that the dimension vector can be defined on $\mcD^b_\mathrm{fd}(A\mathrm{- fgmod})$ to be the alternating sum of the dimensions of the cohomology modules, thus avoiding any issues of infinite dimensionality.}

It is interesting to ask what happens as our central charges leave the upper half plane. This is addressed by Lemma 5.5 of \cite{Bridgeland:2006ss} where we see that it is related to tilting in the derived category. As mentioned in the introduction, Berenstein and Douglas \cite{Berenstein:2002fi} have shown that this is related to Seiberg duality, thus reproducing a standard picture in the physics literature.

Another interesting question is whether this stability condition could be associated to one on the category $\mcD^b(A\mathrm{- fgmod})$. In fact, it is not difficult to see that the central charge map defined above does not factor through $K(A\mathrm{- fgmod})$. To see this, we will use the generalized Ringel resolution of \cite{Bocklandt:2006gc,Ginzburg:2006cy} and discussed in \cite{Bergman:2006gv}. This states that for an arbitrary representation of our CY quiver, $Q$, the following is a projective resolution:
\begin{multline}
0\longrightarrow \hspace{-.4cm} \bigoplus_{i\in \text{Nodes}(Q)}\hspace{-.4cm} P_i \otimes V(i) \longrightarrow \hspace{-.2cm} \bigoplus_{a \in \Arr(Q)}\hspace{-.3cm} P_{s(a)} \otimes V(t(a)) \longrightarrow  \hspace{-.2cm}
\bigoplus_{a \in \Arr(Q)}\hspace{-.3cm} P_{t(a)}\otimes V(s(a)) \longrightarrow\\ \hspace{-.4cm} \bigoplus_{i\in \text{Nodes}(Q)} \hspace{-.4cm} P_{i} \otimes V(i) \longrightarrow
V \longrightarrow 0\ .
\end{multline}
Now, let $d_i = \dim V(i)$. From this resolution, we have that in $K(A\mathrm{- fgmod})$,
\begin{equation}
\label{repcharge}
\begin{split}
[V] &= \sum_{i\in \text{Nodes}(Q)} d_i [P_i] - \sum_{a \in \Arr(Q)} d_{s(a)} [P_{t(a)}] + \sum_{a \in \Arr(Q)}
d_{t(a)} [P_{s(a)}] - \sum_{i\in \text{Nodes}(Q)} d_i [P_i] \\
      &= \sum_{a \in \Arr(Q)}  \left(d_{t(a)} [P_{s(a)}] - d_{s(a)} [P_{t(a)}]\right)\ .
\end{split}
\end{equation}
Applying \eqref{repcharge} to the simple representation $S_i$, we obtain:
\begin{equation}
\label{simpcha}
[S_i] = \sum_{t(a) = i} [P_{s(a)}] - \sum_{s(a) = i} [P_{t(a)}]\ .
\end{equation}

Now, we can rewrite \eqref{repcharge} as:
\begin{equation}
\label{srepcha}
\begin{split}
[V] &= \sum_{a \in \Arr(Q)}  \left(d_{t(a)} [P_{s(a)}] - d_{s(a)} [P_{t(a)}]\right) \\
      &= \sum_{i\in \text{Nodes}(Q)} d_i \left(\sum_{t(a) = i} [P_{s(a)}] - \sum_{s(a) = i} [P_{t(a)}] \right)\\
      &= \sum_{i\in \text{Nodes}(Q)} d_i [S_i]\ .
\end{split}
\end{equation}
In other words, the central charge of any representation can be completely determined by the central charge of the simple representations $S_i$, just as with our assignments. Now, let us choose a skyscraper sheaf. Because $K_X$ is noncompact, the K-theory class of this sheaf is trivial. We also know that is has a dimension vector given by $N_i = \text{rank}(E_i)$. Thus, we have
\begin{equation}
\label{simprel}
\sum N_i [S_i] = [\mcO_x] = 0\ .
\end{equation}
As the central charges of the $[S_i]$ all have nonnegative imaginary part, we see that it is impossible to satisfy this relation. This suggests that the geometric category we should be considering should have some sort of compact support condition on the sheaves. For example, full subcategory of the derived category consisting of objects with proper support\footnote{The support of an object in the derived category is the union of the supports of its cohomology sheaves.} is a Calabi-Yau category (in the sense of Kontsevich). This is currently under investigation.

%We would like to relate these to gauge theory parameters. Let us denote $Z(S_i) = r_i e^{i\pi\xi_i}$. According to Douglas, the mass of the level one string state between nodes $i$ and $j$ is given by \eqref{strmass} as $m^2 = \frac{1}{\alpha'}(\xi_i - \xi_j)$. In the gauge theory, we have the gauge couplings at each node $1/g_i^2$ and the FI-term $f_i$. Computing the D-term potential for a bifundamental, we obtain \textbf{Check!}
%\begin{equation}
%m^2 = g_i^2 {f_i}- g_j^2 f_j\ .
%\end{equation}

%We immediately see that there must be a factor of $\alpha'$ between the FI-terms and the gradings. Thus, let us assume that $\xi_i \ll 1$ for all $i$. Then, $Z(S_i) = r_i (1 + i \pi \xi_i)$. The relation \eqref{fracsum} implies that $\sum r_i = 0$ and $\sum r_i \xi_i$ = 0. Since the diagonal $U(1) \subset \times_i U(N_i)$ in any quiver gauge theory decouples, we can set $\sum N_i g_i^{-2} = 0$ and $\sum f_i N_i = 0$. The $N_i$ is in each expression because we are taking a trace in the $N_i$-dimensional representation, but are only concerned with the $U(1)$ factor. \textbf{Right?} This motivates us to make the assignments:
%\begin{equation}
%\begin{split}
%r_i &= \frac{N_i}{g_i^2} \\
%f_i &= \frac{\xi_i}{\alpha' g_i^2} \ .
%\end{split}
%\end{equation}
%These differ from the assignments of [] by the (possibly conventional) factor of $r_i/ N_i$ in the FI-term.

Next, we would like to see how this notion of stability compares with the notion of stability in a gauge theory. In the construction of the classical moduli space of the quiver gauge theory one does a K\"ahler quotient of the configuration space by the compact gauge group. The FI-terms serve as the moment map in this construction. When they are rational numbers, it is well-known \cite{Kirwan,Luty:1995sd} that this K\"ahler quotient is equivalent to the GIT-quotient by the complexified gauge group. A stability condition in the GIT sense is given by an equivariant line bundle over the configuration space. Since the configuration space is an affine variety, this is just a series of characters for the various gauge groups, $U(N_i)$ whose complexifications are $GL(N_i,\BC)$. Thus, we have a sequence of integers $\theta_i$. Given a set of rational $f_i$ we obtain integral $\theta_i$ by multiplying by the negative of the lcm of the denominators.\footnote{For irrational $f_i$, choose rational numbers sufficiently close so as to not affect the consequent moduli space.} These obey $\sum \theta_i N_i = 0$. We can now use the characterization of semistable and stable representations due to King \cite{King:1994mr}:
\begin{defn}
Let $R$ be a representation of $Q$ with dimension vector $N_i$. Let $\theta_i$ be a vector of integers such that $\sum N_i \theta_i = 0$. Then, we say that the representation is $\theta$-\textbf{semistable} if, for all proper subrepresentations $S\subset R$, $\sum \theta_i \dim(S(i)) \ge 0$. Furthermore, if strict inequality holds for all proper subrepresentations, we say that $R$ is $\theta$-\textbf{stable}.
\end{defn}

To determine stability in Bridgeland sense, note that we are working solely with actual representations (as opposed to complexes of representations). Thus, we can restrict to the t-structure defined by the category of quiver representations. The central charge defines a notion of stability as in definition \ref{stabdef}. In particular, a representation is semi-stable if, for all subrepresentations $U\subset V$, $\xi(U) \le \xi(V)$. As above, we have $Z(S_i) = r_i e^{i\pi\xi_i}$. If we assume that the $\xi_i$ are all close to a given value $\xi_i = \xi + \epsilon_i$, we have
\begin{equation}
Z(U) = \sum_i \dim(U(i)) Z(S_i) \sim e^{i\pi\xi}\sum_i \dim(U(i)) r_i(1 + i \pi \epsilon_i)\ ,
\end{equation}
giving
\begin{equation}
\xi(U) = \xi + \frac{1}{\pi} \tan^{-1} \left(\pi\frac{\sum \dim(U(i)) r_i \epsilon_i}{\sum \dim(U(i)) r_i}\right) \sim
\xi + \frac{\sum \dim(U(i)) r_i \epsilon_i}{\sum \dim(U(i)) r_i}\ .
\end{equation}
Let us denote $\dim(U(i)) = U_i$ and similarly for $V$. The condition $\xi(U) \le \xi(V)$ then becomes:
\begin{equation}
\frac{\sum U_i r_i \epsilon_i}{\sum U_i r_i} - \frac{\sum V_i r_i \epsilon_i}{\sum V_i r_i} \le 0
\end{equation}
which is equivalent to
\begin{equation}
\label{xiineq}
\sum_{i} U_i \left(r_i \left(\epsilon_i \sum_j r_j V_j - \sum_j r_j V_j \epsilon_j\right)\right) \le 0
\end{equation}

Now, fix a dimension vector $N_i$ (for the quiver gauge theory, we have $N_i = \text{rank}(E_i)$). Define
\begin{equation}
\theta_i = -r_i \left(\epsilon_i \sum r_j N_j - \sum r_j N_j \epsilon_j\right)\ .
\end{equation}
and
\begin{equation}
\label{thfdef}
\theta(U) = -\sum_{i} U_i \left(r_i \left(\epsilon_i \sum_j r_j N_j - \sum_j r_j N_j \epsilon_j\right)\right)
\end{equation}
It is straightforward to verify that $\theta(V)  = \sum \theta_i N_i =  0$. In addition, $\theta(U) \ge 0$ imples \eqref{xiineq} by construction. Thus, we have proven\footnote{This result is essentially in Douglas \cite{Douglas:2000ah}.}:

\begin{thm}
Define a stability condition on the derived category of finite-dimensional representation of a Calabi-Yau quiver by a choice of a central charge which only depends on the dimension vector as above. Fix a dimension vector $N_i$ and define $\theta$ as in equation \eqref{thfdef}. Then, for all $\xi_i$ sufficiently close to a fixed angle $\xi$, a representation being (semi-)stable with respect to $\theta$ in the GIT sense implies that it is (semi-)stable with respect to our stability condition on the derived category.
\end{thm}

Finally, we need to relate this to physics in particular to the central charges and gauge couplings of the dual field theory. According to Douglas, the mass of the level one string state between nodes $i$ and $j$ is given by \eqref{strmass} as $m^2 = \frac{1}{\alpha'}(\xi_i - \xi_j)$. In the gauge theory, we have the gauge couplings at each node $1/g_i^2$ and the FI-term $f_i$. Computing the D-term potential for a bifundamental, we obtain
\begin{equation}
m^2 = g_i^2 {f_i}- g_j^2 f_j\ .
\end{equation}

We now identify the $\theta_i$ with the FI-terms, $f_i$, of the gauge theory. In particular, let
\begin{equation}
\begin{split}
\frac{1}{g_i^2} &= r_i \\
f_i &= \frac{r_i \epsilon_i}{\alpha'} - \frac{r_i \sum r_j N_j \epsilon_j}{\alpha'\sum r_j N_j}
\end{split}
\end{equation}
If we change our conventions to consider $f_i/r_i$ as the appropriate term, these only differ from the assignments of \cite{Berenstein:2002fi,Douglas:2000ah} by the $\alpha'$ and an overall constant.

This approach to this issue has closely followed that of Douglas and collaborators. There is another approach, however, due to Aspinwall \cite{Aspinwall:2004mb}. Define a $\theta(U)$ as follows:
\begin{equation}
\theta(U) = - \text{Im }\frac{\sum U_i r_i e^{i\pi\xi_i}}{\sum N_i r_i e^{i\pi\xi_i}}\ .
\end{equation}
It is straightforward to see that this is linear on the dimension vector, obeys $\theta(V) = 0$ and $\theta(U) \ge 0$ implies $\xi(U) < \xi(V)$. This shows that there exists a theta which reproduces Bridgeland stability for a given dimension vector. This also reduces to the above formulae in the case that the gradings almost align. However, the coefficients in this theta do not reproduce the usual expressions for the FI-terms in terms of the gradings.

Regardless which choice of theta we make, it is not the case that we can say that Bridgeland stability and GIT stability are equivalent. For one, the GIT notion of stability is restricted to a fixed dimension vector while that of Bridgeland makes no such assumption. This is probably not a serious issue as we have seen that fixing a dimension vector can be rephrased as fixing to a subspace of the K-theory. A more serious issue is that GIT gives a \textit{moduli space} of stable objects while no one has to my knowledge defined the appropriate notion of a coarse moduli space of Bridgeland stable objects in a triangulated category.

To conclude this section, we will note that any skyscraper sheaf not supported on the zero section of $K_X$ is a Bridgeland stable object in any stability condition as defined in this section. This is a counterpart to the theorem of \cite{Bergman:2005kv} that $K_X$ with the zero section collapsed embeds into the gauge theory moduli space. Thus, a D3-brane anywhere outside the zero section is stable. When it is located on the zero section, however, it becomes part of Bridgeland's category $\mcD_0^b(\mathrm{Coh}(K_X))$. This means that it corresponds to a tiny representation and hence is at best semistable as the only simple representations in this category are the $S_i$. Thus, there exists a decay chain into a collection of the $S_i$. On dimensional grounds, we know that the number of these must exactly replicate the $N_i$, the dimension vector of the original representation. Thus, we have verified the existence of the decay of the D3-brane into fractional branes precisely when it is located on the zero section of $K_X$, and it is stable everywhere else.

In order to show stability of $\mcO_x$ off the zero section, it suffices to prove the following:

\begin{thm}\label{th:sta} Let $\mcO_x$ be a skyscraper sheaf not supported on the zero section of $K_X$. Then, the representation corresponding to $\mcO_x$ under the equivalence of categories given in section \ref{sec:cat} is a simple representation.
\end{thm}

The proof of this theorem is given in the next section. In combination with the results of \cite{Bergman:2005mz}, it shows that the complement of the zero section embeds into the moduli space of representations (in the GIT sense) with any GIT stability condition, and that this embedding is an isomorphism on tangent spaces. In particular, this result does not depend on the exceptional collection being composed of line bundles. It would be interesting to understand what occurs at zero section in this general context.

The relation of this mathematical result to the physics deserves further explanation. What it says is that any D3-brane off the zero section of $K_X$ is stable in any stability condition defined in this section. What can vary, however, is the set of fractional branes that the D3-brane can decay into when located on the zero section. We have seen that the decay is determined by the choice of stability condition which, in turn, is determined by the choice of a category of quiver representations. In other words, the different quivers corresponding to different exceptional collections correspond to different open regions in a generalized K\"ahler moduli space of our theory. Furthermore, quivers related by mutation can arise by passing to adjacent regions in this moduli space, a procedure related to Seiberg duality \cite{Herzog:2004qw,Aspinwall:2004vm,Cachazo:2001sg}. Without further understanding the relation between Bridgeland's space of stability conditions and the physical moduli space, however, we cannot say if all these regions occur in the actual string theory.

\section{The proof of the theorem}

In order to prove theorem \ref{th:sta}, we first need to define the center of a category.

\begin{defn} The center of a category, $Z(\mcC)$, is given by the set of natural transformations from the identity to itself. \end{defn}

For the categories we are using, this is the zeroth Hochschild cohomology group $\HH^0(\mcC)$. As discussed in \cite{Bergman:2006gv}, when $\mcC \cong \mcD^b(Coh(X))$, $\HH^0(\mcC) = \Gamma(\mcO_X)$, \ie global functions on $X$. On the other hand, when $\mcC \cong \mcD^b(A\mathrm{- fgmod})$, we have that $\HH^0(\mcD^b(A \mathrm{- fgmod})) = Z(A)$, the center of the algebra $A$.

An object in the center gives rise to an endomorphism of every object of our category satisfying certain consistency rules. As both $\mcD^b(A\mathrm{- fdmod})$ and $\mcD^b_\mathrm{fd}(A\mathrm{- fgmod})$ are full subcategories of $\mcD^b(A\mathrm{- fgmod})$, the action of the center of $\mcD^b(A\mathrm{- fgmod}) = Z(A)$ descends to an action on them. This action is simple to determine. Given a representation $V$ of $A$, we have a map $r : A \to \End(V)$. Given an element $z \in Z(A)$, we have a map $r(z) : V \to V$. Essentially by definition, this is an endomorphism of the representation and gives rise to an endomorphism in the derived category.

The action of the center on the derived category of coherent sheaves is also straightforward. Any object in this category can be represented as a bounded complex of coherent sheaves, \ie $\mcO_X$-modules. Thus, they are certainly acted on by global functions, and it is obvious that this gives rise to an endomorphism in the derived category.

Now, recall our situation. We have an equivalence of categories $F : \mcD^b(Coh(Y)) \to \mcD^b(A \mathrm{- fgmod})$ given by a locally free tilting sheaf $T$ with $A = \End_Y(T)^\mathrm{op}$. In particular, $F(\mcE) = \mathbb{R}\Hom(T,\mcE)$ which is a complex of $\End_Y(T)$-modules. We need the following lemmas. In what follows, ``point" will always refer to a closed point.

\begin{lma} \label{lma:stalk} Let $\mcA$ be a non-zero indecomposable object in $\mcD^b(Coh(Y))$ such that the support of its cohomology sheaves is solely at the point $p$. Then $\mcA$ is isomorphic to the image of a shift of the skyscraper sheaf $\mcO_p$ in the derived category. \end{lma}

\begin{proof} Given any sheaf $\mcF$ and point $p$, there is a map $\mcF \to \mcF_p / \mathfrak{m}_p \mcF$ where the latter sheaf is a skyscraper supported at the point $p$ whose fiber is the fiber of $\mcF$ at $p$. Thus, if we represent $\mcA$ as a bounded complex of sheaves, there is a chain map from $\mcA$ to a complex of skyscrapers supported at the point $p$. This obviously induces an isomorphism in cohomology, so it is an isomorphism in the derived category. Since coherent sheaves on a point are vector spaces, the assumption of indecomposability means that we must have only a one-dimensional vector space, thus proving the lemma. \end{proof}

\begin{lma} \label{lma:sky} $F(\mcO_p)$ can be considered as the image in any of the derived categories of an object in $A \mathrm{- fdmod}$, and this object is given by the dual of the fiber of $T$ at the point $p$ with its action of $\End_Y(T)$. Furthermore, the action of the center is given by scalar multiplication by the value of the global function at the point $p$.\end{lma}

\begin{proof} This is essentially obvious. Since skyscraper sheaves have no higher cohomology, $\mathbb{R}\Hom(T,\mcO_p)$ is an honest representation. Furthermore, the vector space $\Hom(T,\mcO_p)$ is precisely the dual of the fiber of $T$ at the point $p$. It is finite dimensional by the coherence of $T$. The action of the center is the action of global functions on the fiber and is precisely as given in the statement of the lemma. \end{proof}

We can now prove theorem \ref{th:sta} which we restate in a more general form.

\begin{thm} Let $Y$ be a smooth variety with a locally free tilting sheaf $T$,  $\pi$ be the projection from $Y$ to its affinization, and $p$ be a point in $Y$ such that $\pi^{-1}\pi(p) = p$. Then the $\End_Y(T)$ module corresponding to $\mcO_p$ is  a simple module.\end{thm}

\begin{proof} By Lemma \ref{lma:sky}, the representations corresponding to skyscraper sheaves are acted upon by the center, $Z$, by scalar multiplication. Let $M$ be an indecomposable subrepresentation of $F(\mcO_p)$. We want to show that $M$ is isomorphic to $F(\mcO_p)$. $Z$ acts on $M$ by precisely the same scalar multiplication as it acts on $F(\mcO_p)$. By the equivalence of categories, there exists an object $\widetilde{M}$ in the derived category of coherent sheaves such that $F(\widetilde{M}) \cong M$. The center acts on the cohomology sheaves of this object by scalar multiplication. This scalar multiplication provides a character of the ring of global functions and thus a point, $q$, on the affinization of $Y$. By examining the action of the center on the stalks, we see that the support of the cohomology sheaves of $\widetilde{M}$ lies in the inverse image of $q$ in $Y$. Furthermore, by construction, the point $p$ projects to $q$ on the affinization. As we have assumed that $\pi^{-1}\pi(p)=p$ , we see that the cohomology sheaves of $\widetilde{M}$ are supported at $p$. Thus, by lemma \ref{lma:stalk}, $\widetilde{M}$ is a shift of a skyscraper sheaf. Since the skyscraper corresponds to an actual representation, \ie an object in the heart of the t-structure corresponding to the abelian category of representations, we see that shifting it would take it out of the heart. As we have assumed that $M$ is an honest representation and lies in the heart, it must be isomorphic to $F(\mcO_p)$. \end{proof}

In our case, we have $Y = K_X$, and the failure of the argument for a skyscraper supported on the zero section is straightforward. The inverse image of the corresponding point on the affinization is the entire zero section of $K_X$. The simple tiny representations are push-forwards of sheaves on $X$ by the zero section and are thus possible subrepresentations. The derived category of coherent sheaves whose cohomology sheaves are supported on the zero section is precisely the category $\mcD_0^b(A\mathrm{- fgmod})$ considered by Bridgeland whose heart is a finite length Abelian category.

\acknowledgments

I would like to thank David Ben-Zvi, Tom Bridgeland and Jason Kumar for useful conversations and e-mails on this project. I would also like to thank the Institute for Advanced Study for their hospitality while a portion of this work was being completed. This material is based upon work supported by the 
National Science Foundation under Grant Nos. PHY-0505757 and PHY-0555575 and Texas A\&M University.

\bibliographystyle{utphys}
\bibliography{thebib}
\end{document}